\documentclass[useAMS,usenatbib,usegraphicx]{mn2e}
\usepackage[draft]{hyperref} 
\usepackage{xfrac} 
\usepackage[svgnames]{xcolor}  %

\newcommand{\arcdeg}{\degr}
\newcommand{\uv}{\mbox{$u$-$v$}}

\newcommand{\Msol}{\mbox{$M_{\sun}$}}
\newcommand{\Msolxyr}{\mbox{$M_{\sun}$~yr$^{-1} / 10$~km~s$^{-1}$}}

\newcommand{\kms}{\mbox{km s$^{-1}$}}

\newcommand{\muasyr}{\mbox{$\mu$as~yr$^{-1}$}}

\newcommand{\Jb}{\mbox Jy~beam$^{-1}$}
\newcommand{\muJb}{\mbox{$\mu$Jy~beam$^{-1}$}}

\newcommand{\spost}{_{\rm post} }
\newcommand{\simpact}{_{\rm impact} }
\newcommand{\rx}{\mbox{$r_{\rm 16}$}}

\newcommand{\lesssim}{\mbox{\raisebox{-0.3em}{$\stackrel{\textstyle <}{\sim}$}}}
\newcommand{\gtrsim}{\mbox{\raisebox{-0.3em}{$\stackrel{\textstyle >}{\sim}$}}}
\newcommand{\tablenotemark}[1]{$^{\mathrm #1}$}
\newcommand{\tablenotetext}[2]{\noindent$^{\mathrm #1}$ #2\\}
\newcommand{\phn}{\phantom{1}}

\newcommand{\slugcomment}[1]{\date{#1}}

\newcommand{\araa}{Ann.\ Rev.\ Astron.\ Astrophys.}
\newcommand{\aap}{Astron.\ Astrophys.}

\newcommand{\apj}{ApJ}
\newcommand{\apjl}{ApJL}
\newcommand{\apjs}{ApJS}

\newcommand{\mnras}{MNRAS}

\newcommand{\pasa}{PASA}

\title{SN 2014C: VLBI Images of a Supernova Interacting with a
  Circumstellar Shell}

\author[Bietenholz et al]{Michael F. Bietenholz$^{1,2}$,
Atish Kamble$^3$,
Raffaella Margutti$^{4}$, 
Danny Milisavljevic$^3$,  \and 
Alicia Soderberg$^3$
\\
$^1$Hartebeesthoek Radio Observatory, PO Box 443, Krugersdorp,
1740, South Africa \\
$^2$Department of Physics and Astronomy, York University, Toronto,
M3J~1P3, Ontario, Canada \\
$^3$Harvard-Smithsonian Center for Astrophysics, 60
Garden Street, Cambridge, MA 02138, USA \\
$^4$Center for Interdisciplinary Exploration and Research in Astrophysics (CIERA) and Department of Physics and Astronomy,\\
Northwestern University, Evanston, IL 60208, USA \\
}

\begin{document}

\slugcomment{Accepted to {\em MNRAS};  \today}

\pagerange{\pageref
{firstpage}--\pageref{lastpage}} \pubyear{2017}
\maketitle
\label{firstpage}


\begin{abstract}
  
We report on VLBI measurements of supernova 2014C at several epochs
between $t = 384$ and 1057 days after the explosion.  SN~2014C was an
unusual supernova that initially had Type Ib optical spectrum, but
after $t = 130$~d it developed a Type IIn spectrum with prominent
H$\alpha$ lines, suggesting the onset of strong circumstellar
interaction.  Our first VLBI observation was at $t = 384$~d, and we
find that the outer radius of SN~2014C was $(6.40 \pm 0.26) \times
10^{16}$~cm (for a distance of 15.1 Mpc), implying an average
expansion velocity of $19300 \pm 790$~\kms\ up to that time. At our
last epoch, SN~2014C was moderately resolved and shows an
approximately circular outline but with an enhancement of the
brightness on the W side.  The outer radius of the radio emission at
$t = 1057$~d is $(14.9 \pm 0.6) \times 10^{16}$~cm.  We find that the
expansion between $t = 384$ and 1057~d is well described by a constant
velocity expansion with $v = 13600 \pm 650$~\kms.  SN~2014C had
clearly been substantially decelerated by $t = 384$~d.  Our
measurements are compatible with a scenario where the expanding shock
impacted upon a shell of dense circumstellar material during the first
year, as suggested by the observations at other wavelengths, but had
progressed through the dense shell by the time of the VLBI
observations.
\end{abstract}


\section{Introduction}
\label{sintro}

A complex picture has been emerging of the ends of the lives of
massive stars, and how they shed matter in stellar winds before
undergoing a core-collapse supernova (SN) explosion. It is often
assumed that the stellar winds are relatively steady, leading to a
circumstellar medium (CSM) with density, $\rho \propto r^{-2}$
\citep[e.g.][]{Weaver+1977, Chevalier1982b, Dwarkadas2005,
  ChevalierI2011}.  However, recently evidence has been mounting that
many massive stars experience highly variable stellar winds and
eruptive mass-loss in the period before the explosion \citep[see][for
  a review]{Smith2014}.
The stellar winds are driven from the surface of the star, whereas the
processes leading up to the core collapse occur deep in the interior,
so it is unclear what drives mass-loss events shortly before the SN
explosions.

In the case of Type Ib/c SNe, the so-called ``stripped-envelope'' SNe,
which arise from stars that have lost much of their H envelopes prior
to the explosion, evidence for non-steady mass loss shortly prior to
the explosion has been seen in several SNe.  In particular, modulated
radio lightcurves have been seen in several Type I b/c SNe
\citep{Soderberg+2006e, WellonsSC2012, Milisavljevic+2013a}, which
suggest modulations in the CSM density as a function of radius and
therefore temporal variation of the mass-loss in the period preceding
the SN explosion.

SN~2014C is a particularly interesting case.  It was discovered on
2014 January 5 \citep{Kim+2014} in the nearby early-type spiral galaxy
NGC~7331 by the Lick Observatory Supernova Search.  We adopt the
updated Cepheid distance of $D = 15.1 \pm 0.7$~Mpc from
\citet{Saha+2006}.
We also adopt an explosion date, $t = 0$, of 2013 December 30.0 (UT) =
MJD 56656.0, as determined by \citet{Margutti+2017_SN2014C} from
bolometric lightcurve modelling.

At the time of its discovery, SN~2014C's spectrum was that of an
ordinary, H-stripped Type Ib supernova
\citep{Kim+2014,Tartaglia+2014}, but over the next year it evolved
into a Type IIn spectrum showing prominent H$\alpha$-lines indicating
strong CSM interaction \citep{Milisavljevic+2015_SN2014C,
  Margutti+2017_SN2014C}.  It also, unusually, kept brightening in
X-rays till $t \sim 1$~yr \citep{Margutti+2017_SN2014C}, and H$\alpha$
emission was still detected as late as $t \simeq 892$~d
\citep{Vinko+2017}.
In the mid-infrared, it had an almost constant brightness till $t \sim
800$~d, with even a possible slight re-brightening near $t \sim 250$~d
\citep{Tinyanont+2016}.

It was detected early on in radio, both at 7 GHz by the Very Large
Array \citep{Kamble+2014b}, and at 85~GHz by Combined Array for
Research in Millimeter Astronomy \citep{Zauderer+2014}.  A 16-GHz
radio lightcurve from the Arcminute Microkelvin Imager was presented
by \citet{Anderson+2017}, who found a very unusual two-peaked
lightcurve, with the first peak at $t \simeq 80$~d, and a second at $t
\simeq 400$~d. A paper on our multi-frequency flux-density monitoring
of SN~2014C with the Jansky Very Large Array is in preparation (Kamble
et al.).

\citet{Milisavljevic+2015_SN2014C} and \citet{Margutti+2017_SN2014C}
suggest a scenario where SN~2014C's progenitor exploded inside a
relatively low-density cavity, but within a year, the expanding shock
encountered a massive H-rich shell (probably $\sim 1$~\Msol) at
radius, $\rx \sim 6$, where \rx\ is a dimensionless radius, and $\rx =
r/(10^{16} \; {\rm cm})$.  Such a shell is not expected from the
standard metallicity-dependent line-driven mass-loss scenario
\citep{KudritzkiP2000, NugisL2000}, and requires a different, highly
time-dependent mass-loss mechanism that is active during the last
centuries before the explosion.

A key ingredient to understanding SN~2014C is knowing basic
parameters, in particular the (time-dependent) radius of the expanding
ejecta and the corresponding expansion speed.  Very contradictory
values have been proposed: \citet{Anderson+2017} suggest values of
\rx\ of $0.332 \pm 0.007$, based on modelling the lightcurve bumps on
the assumption that the dominant absorption mechanism is synchrotron
self-absorption (SSA), although they do warn that this assumption may
not be appropriate for SN~2014C\@.
The size reported by Anderson et al.\ would imply relatively low
average expansion velocities of $\lesssim 5000$~\kms\ for $t = 80$~d,
whereas typically Type Ib SNe have velocities several times higher
\citep[e.g.][]{Chevalier2007, SNIbc-VLBI}.  As mentioned, there is
considerable evidence of SN~2014C's shock impacting on a dense shell
of circumstellar material (CSM) between $t = 30 \sim 130$~d after the
explosion, which would likely entail high shock velocities in the
period before the impact.  For example, an average speed\footnote{Note
  that \citet{Margutti+2017_SN2014C} give a value of 44000~\kms\ for
  the instantaneous velocity of the forward shock when it impacts on
  the shell; our value, which is merely the radius divided by the
  time, is the {\em average} value between the time of the explosion
  and the time of impact, and is slightly higher than the
  instantaneous value because the shock is decelerating.}  of
49000~\kms, which is an order of magnitude higher than the value
implied by Anderson et al., ensues from
\citet{Margutti+2017_SN2014C}'s particular values for the time of the
shock's impact on the shell, 130~d, and its radius, $\rx \sim 5.5$.

VLBI observations are the only way to resolve the forward shock and to
therefore obtain direct observational constraints on its size, and
thus also on the shock velocity.  We therefore undertook VLBI
observations of SN~2014C with the goals of determining the radius at
various times.

The VLBI observations also hold the promise of revealing the
morphology of the radio emission.  There are only a handful of SNe
where the radio emission can be resolved \citep[see][for recent
  reviews]{SNVLBI_Crete, SNVLBI_Cagliari}, so given SN~2014C's close
distance and bright radio emission, it held the promise of adding to
our catalogue of SNe with resolved emission.

\section{Observations and Data Reduction}
\label{sobs}

We obtained four VLBI observing sessions on SN~2014C with the National
Radio Astronomy Observatory (NRAO)\footnote{The National Radio
  Astronomy Observatory, NRAO, is a facility of the National Science
  Foundation operated under cooperative agreement by Associated
  Universities, Inc.}
High Sensitivity Array (HSA) between 2015 January and 2016 April. We
give the particulars of the observing runs in Table~\ref{tobs}.

The HSA includes the Robert C. Byrd telescope at Green Bank, GB
($\sim$105~m diameter).  At the time of our observations, GB was
affected by a bug, that caused GB phases to reset at the start of each
scan, and therefore rendered phase-referencing impossible for GB
data. The GB data could be recovered by self-calibrating in phase on a
per-scan basis if the signal-to-noise was high enough. In the cases
where this was possible we did so, and our images and model-fitting
results are based on this self-calibrated data.  Where such
self-calibration was not possible we excluded the GB data.  All our
astrometric results are based on un-self-calibrated data excluding GB.

\begin{table*}
\begin{minipage}[t]{\textwidth}
\caption{VLBI Observations of SN2014C}
\label{tobs}
\begin{tabular}{l c l c c@{ }c@{\,}c@{\,}c@{\,}l}
\hline
Date &  Proposal & Telescopes\tablenotemark{b} &MJD\tablenotemark{c}~ &
Freq. & Total flux & Peak\tablenotemark{d} & Image rms & Convolving beam  \\
 & code\tablenotemark{a} &  &  &  & density\tablenotemark{e} & Brightness &  & ~~~~FWHM \\
& & & & (GHz) & (mJy)  & (m\Jb) & (\muJb) & (mas $\times$ mas, \arcdeg) \\
\hline
\hline
2015 Jan 17 & 14B-500 & VLBA, GB & 57039.8 & 22.1  & 17.8 & \phn9.0& 103~ &  $0.68 \times 0.33, 12\arcdeg$ \\ 
2015 Jan 17 & 14B-500 & VLBA, GB & 57039.8 &\phn8.4& 25.2 & 21.7 & 80 & $1.55 \times 0.80, -3\arcdeg$ \\ 
2015 Oct 05 & 15B-312 & VLBA, GB\tablenotemark{f} 
                                & 57301.2 & 22.1  & \multicolumn{4}{l}{no image made due to GB phasing problems} \\
2015 Oct 05 & 15B-312 & VLBA, GB\tablenotemark{{f,g}} 
                                        & 57301.2 &\phn8.4& 26.4 & 20.0  & 80 & $1.55 \times 0.80, -3\arcdeg$ \\ 
2016 Apr 30 & 15B-312 & VLBA, GB, Y, EB & 57508.5 &\phn8.4& 23.0 & 15.6 & 27 & $1.55 \times 0.80, -3\arcdeg$ \\  
2016 Nov 20 & 15B-312 & VLBA, GB, Y, EB & 57713.0 &\phn8.4& 22.8 &\phn8.8& 21& $1.32 \times 0.44, -12\arcdeg$ \\ 
\hline
\end{tabular} 
\\
\tablenotetext{a}{NRAO observing code}
\tablenotetext{b}{VLBA = NRAO Very Long Baseline Array, $10 \times
  25$~m diameter; GB = Robert C. Byrd telescope at Green Bank,
  $\sim$105~m diameter; Y = the Jansky Very Large Array in
  phased-array mode, equivalent diameter 94~m; EB = the Effelsberg
  antenna, 100~m diameter.}
\tablenotetext{c}{Modified Julian Date of midpoint of observation}
\tablenotetext{d}{Total (CLEANed) flux density in image}
\tablenotetext{e}{The peak brightness in the image}
\tablenotetext{f}{EB and Y were scheduled, but the observations failed}
\tablenotetext{g}{The image was super-resolved by $\sim$6\% to match
  the resolution of other images}
\end{minipage}
\end{table*}

We observed at both 8.4 and 22.1~GHz, recording both senses of
circular polarization over a bandwidth of 256~MHz.  As usual, a
hydrogen maser was used as a time and frequency standard at each
telescope, and we recorded with the RDBE/Mark5C wide-band system at a
sample-rate of 2~Gbps, and correlated the data with NRAO's VLBA DiFX
correlator.  All the observing runs were 8~h in length, with the time
was divided approximately equally between observations at 22 and 8.4
GHz for the two runs in 2015, while in 2016 we observed only at
8.4~GHz.

We phase-referenced our observations to the sources VCS3 J2235+3418,
0.24\arcdeg\ away from SN~2014C, and VCS1 J2248+3718,
2.90\arcdeg\ away.  We will refer to the two just as J2235+3418 and
J2248+3718, respectively.  We used J2235+3418 as a primary reference
source at 8.4 GHz, as it is closer on the sky.  However, at 22~GHz it
is too weak for reliable phase-referencing, so we used the stronger
but somewhat more distant J2248+3718\@.  We included some scans of
J2235+3418 at 22 GHz and of J2248+3718 at 8.4 GHz.  This served two
purposes: firstly it allows us to align the images at the two
frequencies accurately, and secondly allows us to check for any
possible proper motion of the reference sources.

The data reduction was carried out with NRAO's Astronomical Image
Processing System (AIPS).  The initial flux density calibration was
done through measurements of the system temperature at each telescope,
and improved through self-calibration of the phase-reference sources.

\section{VLBI images}  
\label{svlbiimg}

\begin{figure}
\centering
\includegraphics[width=0.93\linewidth]{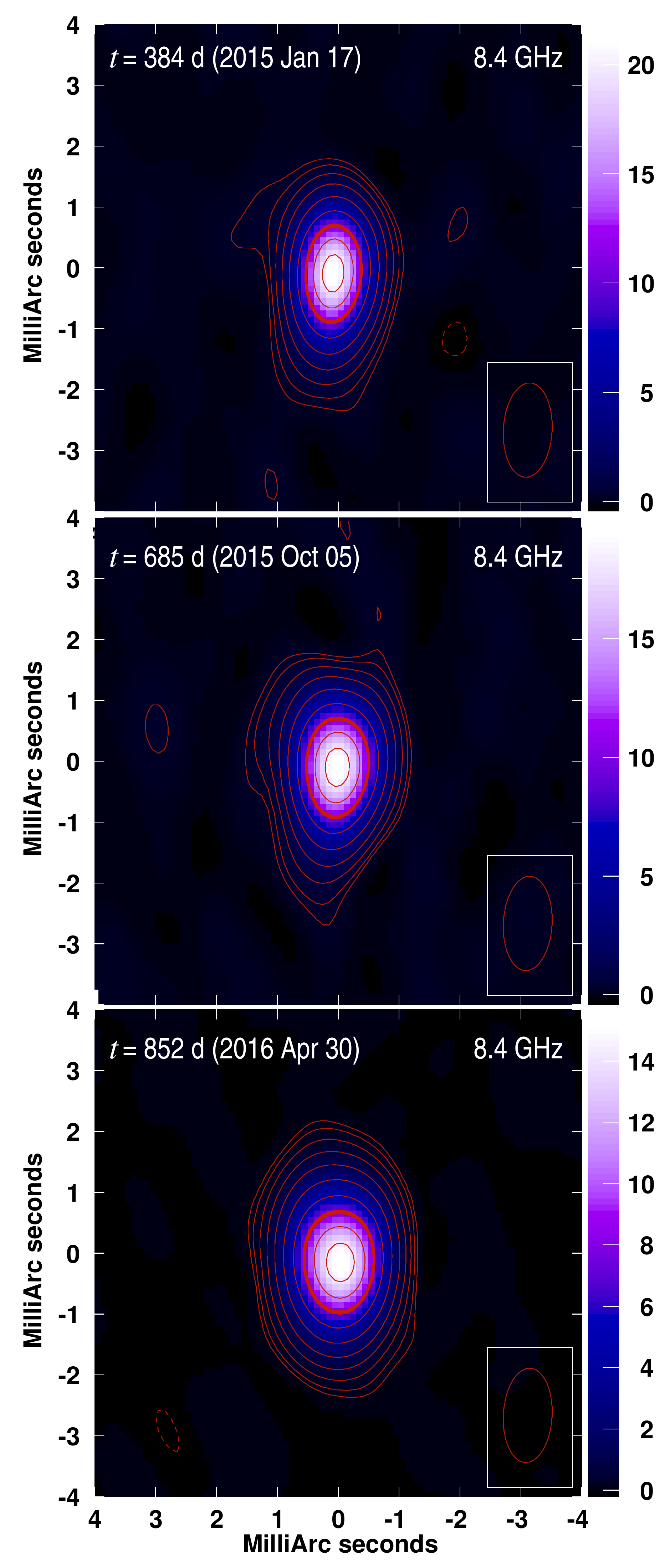}
\caption{The 8.4-GHz VLBI images from 2015 January 17, 2015 October 05
  and 2016 April 30, all rendered at a common FWHM resolution of $1.55
  \times 0.8$~mas at p.a.\ $-3$\arcdeg, indicated at lower left in
  each pane.  Both the contours and the colorscale show the image
  brightness, the latter labelled in m\Jb.  The total CLEANed flux
  densities, peak brightnesses, rms backgrounds and CLEAN beams are
  given in Table~\ref{tobs}.  For 2015 January 17 and 2015 October 5,
  the contours are at $-1$, 1, 2, 4, 8, 16, 30, {\bf 50} (emphasized),
  70 and 90\%, of the peak brightness.  The 2015 October 5 image was
  super-resolved by $\sim$6\% to match the resolution of the others.
  For 2016 April 30, the contours are as for the previous two except
  the lowest contours are at $-0.6$, 0.6 and 1\%.}
\label{fimgseq}
\end{figure}

\begin{figure}
\centering
\includegraphics[width=0.98\linewidth]{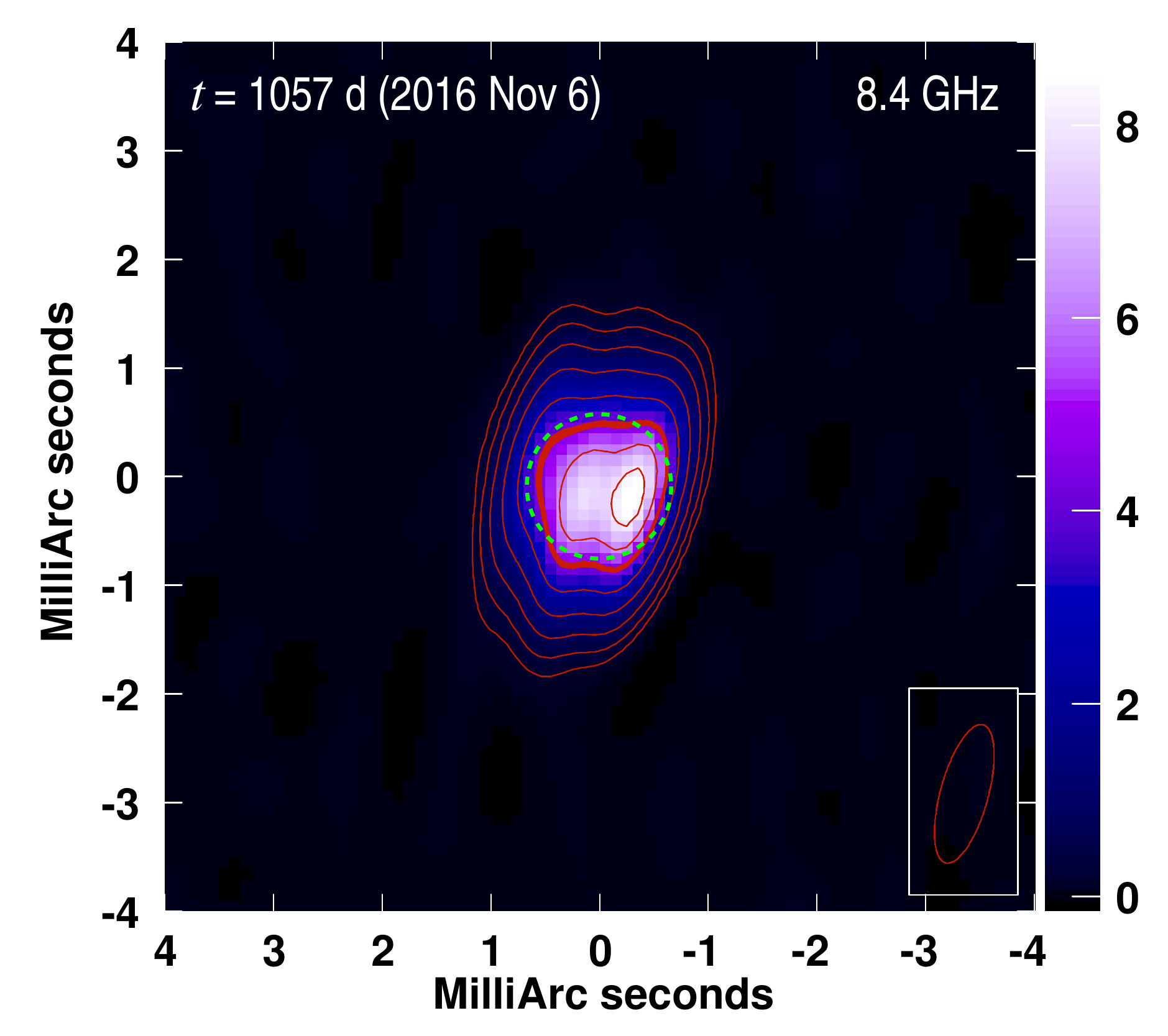}
\caption{The 8.4-GHz VLBI Image from 2016 November 6 or $t = 1057$~d.
  The total CLEANed flux density was 22.8 mJy, and the rms of the
  background brightness was 21 \muJb.  Both the contours and the
  colour show the image brightness, with the colorscale being labelled
  in mJy~beam$^{-1}$.  The contours, in red, are at $-1$, 1, 2, 4, 8,
  16, 30, {\bf 50} (emphasized), 70 and 90\% of the peak brightness,
  which was 8.81 m\Jb.  The FWHM resolution of $1.32 \times 0.44$~mas
  at p.a.\ $-15$\arcdeg\ is indicated at lower left.  The dashed
  (green) circle shows the outer radius of a fitted spherical shell
  model (see text, \S \ref{ssize}).}
\label{fimglast}
\end{figure}

We show a sequence of the first three 8.4-GHz VLBI images of SN~2014C
from 2015 January 17 to 2016 April 30 ($t = 384 - 852$~d) in
Figure~\ref{fimgseq}, and the 8.4-GHz image from 2016 November 6 ($t =
1057$~d) in Figure \ref{fimglast}.  The total flux densities, peak
brightnesses, and background rms brightness values are given in
Table~\ref{tobs}.  The images were deconvolved using the CLEAN
algorithm, with AIPS robustness parameter set to $-4$ which gives a
weighting close to uniform.  To increase the reliability of the images
we used the square root of the data weights in the imaging, which
results in more robust images less dominated by a few very sensitive
baselines in the HSA.

For our last image (2016 November 6, Figure~\ref{fimglast}), the total
CLEANed flux density was 22.8 mJy, the peak brightness was 8.81 m\Jb,
and the rms of the background brightness was 21 \muJb.  The FWHM
resolution or CLEAN beam was somewhat elongated due to the low
declination of the source, and was of $1.32 \times 0.44$~mas at
p.a.\ $-15$\arcdeg.  At this epoch, SN~2014C is moderately resolved in
an approximately E-W direction with the E-W extent of the source
(estimated by the 50\% contour) being $\sim$3 times the FWHM
resolution.  In the perpendicular direction, the source diameter is
comparable to the resolution.

Although the lower brightness contours of SN~2014C in
Figure~\ref{fimglast} are relatively elongated, this is merely due to
the convolution with the elongated CLEAN beam. The 50\% contour of the
image, which would correspond approximately to the perimeter of a
hard-edged source, is more circular, suggesting that the true source
shape is likely not greatly elongated.  There is a distinct brightness
enhancement on the W side of the source.  Since the dynamic range of
this image is quite high, this enhancement is almost certainly real as
it is much larger than either the off-source noise or any expected
deconvolution artefacts.

\section{Size, Expansion Speed and Proper Motion}
\label{ssize}

Although SN~2014C is moderately resolved in our last VLBI image, it is
less so in our earlier images, and we turn to fitting geometrical
models in the Fourier transform or \uv~plane to most accurately
determine the size evolution. \citet{SN93J_Manchester} showed that in
the case of SN~1993J, the results obtained through \uv~plane
modelfitting are superior to those obtained in the image plane.  We
again used the square root of the data weights in the fitting, which
makes the results more robust at the expense of some statistical
efficiency.  We used the AIPS task OMFIT to fit the models.

The expected structure of the radio emission of a SN is that of a
spherical shell, with the radio emission arising between the forward
shock which is driven into the CSM and a reverse shock that is driven
back into the freely-expanding ejecta.  Indeed, the few SNe that have
been reasonably resolved mostly show a morphology approximately
consistent with a projected spherical shell, albeit usually with some
localized brightness enhancements along the ridge of the projected
shell \citep[see, e.g.][]{SNVLBI_Cagliari}.  Our last image of
SN~2014C (Fig. \ref{fimglast}) indeed also shows a structure that can
also be interpreted within this context, although it does show a
distinct departure from circular symmetry in the presence of a
brightness enhancement on the W side.

If the supernova is optically thick, then the expected morphology is
closer to a uniform disk. This was the case for our
2015 Jan.\ epoch at 8.4 GHz, as can be seen from the SED in
\citet{Margutti+2017_SN2014C}, which shows that the turnover frequency
was around 20~GHz at that time.
We did fit both a disk and a spherical shell model to the 2015
Jan.\ data and found that indeed the disk model fitted better for the
8.4-GHz data, at which frequency the SN was still optically thick,
while the shell model fitted better at 22 GHz, where the SN was
optically thin.  However, the fractional difference in $\chi^2$
between the two models was very small ($<10^{-3}$) so we do not
consider the difference significant.  The fitted outer radii are
expected to be very similar for the two models, and indeed our fitted
outer radius values agreed to within half the uncertainties.  For the
sake of consistency, were therefore use only the values from the
spherical shell model below, since those should be more appropriate
for most epochs and frequencies, and as we have shown the differences
are small even if the SN is still optically thick.

The data do not reliably determine the thickness of the spherical
shell even at our last epoch when the SN is most resolved, so we
assume a thickness of 20\% of the outer radius, which has been shown
to be appropriate in the case of SN~1993J \citep{SN93J-3, SN93J-4},
and is near the value expected on theoretical grounds in the case of a
simple CSM structure \citep{Chevalier1982a, JunN1996a}.
Our fitted outer radius is only weakly
dependent on the assumed shell thickness, and reasonable deviations
from the assumed value of 20\% will not change our outer radii by more
than our stated uncertainties.

We use the outer radius of this fitted spherical shell model to
estimate the outer radius of SN~2014C\@.  The purely statistical
uncertainties on the fitted sizes were small in all cases ($\lesssim
0.010$~mas).  Given the approximate nature of the shell model and the
fact that, because of the antenna-dependent gain calibration, the
calibrated visibility measurements are not statistically independent,
systematic contributions will dominate the uncertainty in the fitted
sizes.  We estimated two different contributions to the systematic
uncertainties, which we add in quadrature to the statistical one for
our total uncertainties.

The first contribution was estimated using jackknife resampling
\citep{McIntosh2016}.  Specifically, we dropped the data from each of
the antennas in the VLBI array in turn and calculated $N_{\rm
  antenna}$ new estimates of the fitted size, and the scatter over
these $N_{\rm antenna}$ values allows one to estimate the uncertainty
of the original value which included all antennas.  Since the
uncertainty should scale with the resolution, we compared the results
to the FWHM resolution, for which we take the geometric mean of the
major and minor axes of the fitted restoring beam.  We carried out this
procedure for four epochs, 2015 Jan.\ 17, 22~GHz; 2015 Oct.\ 05,
8.4~GHz, 2016 Apr.\ 30, 8.4~GHz and 2016 Nov.\ 20, 8.4~GHz, and
obtained estimates of the uncertainty of 2.1, 1.2, and 1.2 and 4.2\%
of FWHM resolution for the three epochs respectively.  Based on this
test, we took a rounded 2\% of the geometric mean of the major and
minor axes of the restoring beam as the uncertainty for all epochs.

The second contribution is an estimate of the effect of any gain
mis-calibration on the fitted sizes. For marginally resolved sources,
such as SN~2014C, the fitted size is correlated with the antenna
gains, which are not exactly known.  We estimated this contribution to
the uncertainty in a Monte-Carlo fashion by repeatedly randomly
varying the individual antenna gains by 10\% (rms), and then
re-fitting the spherical shell models.  This estimate should be
conservative as it is unlikely that our antenna gains would be wrong
by as much as 10\%.

We give the fitted outer radii and the total uncertainties in
Table~\ref{tradii}, and plot them in Figure~\ref{fexpand}.  For our
last epoch ($t = 1057$~d), we also plot a circle showing the outer
radius of the fitted model in Figure~\ref{fimglast} to show the
relation between the fitted model and the VLBI image.

Given the brightness enhancement to the W seen in the $t = 1057$~d
image, the circularly symmetric spherical shell model is clearly only
an approximate description of SN~2014C\@.  However, barring any gross
changes in the morphology as it expands, our fitted radii should give
a reliable picture of the expansion of the SN.

For the first two epochs, we had observations at both 22.1 and
8.4~GHz.  In both cases we found that the radii measured at 8.4~GHz
were $\sim$20\% larger than the corresponding 22-GHz ones, however,
the difference is only $\sim 1\sigma$ in each case.  Absorption is
strongly frequency dependent, and so might cause a variation with
frequency of the apparent size.  If SN~2014C were optically thick
still at 8.4~GHz at these epochs ($t = 383, 645$~d), we would expect
real brightness distribution to be approximately disk-like.  If this
case, our fit using a spherical shell model would slightly
{\em under}estimate the true radius \citep[for a discussion of this issue in
  the case of SN~1993J, see][]{SN93J-2}.  However, the lower frequency
should be more affected than the higher one, so in this case one would
expect the fitted radii at the lower frequency to be too small, which
is the opposite of what we find.  We therefore consider it unlikely
that absorption effects cause the differences of radius with frequency
we find, which as mentioned, are of questionable significance.

\begin{table}
\begin{minipage}[t]{0.48\textwidth}
\caption{Radius Measurements}
\label{tradii}
\begin{tabular}{l c r c c}
\hline
Date & MJD & Age & Frequency & Outer Radius\tablenotemark{a} \\
     &     & (d) &  (GHz)    & (mas) \\
\hline
\hline
2015 Jan 17 & 57039.8 & 384  & 22.1    & $0.280 \pm 0.012$ \\ 
2015 Jan 17 & 57039.8 & 384  & \phn8.4 & $0.330 \pm 0.045$ \\ 
2015 Oct 05 & 57301.2 & 645  & 22.1    & $0.359 \pm 0.085$ \\ 
2015 Oct 05 & 57301.2 & 645  & \phn8.4 & $0.439 \pm 0.034$ \\ 
2016 Apr 30 & 57508.5 & 852  & \phn8.4 & $0.524 \pm 0.020$ \\ 
2016 Nov 20 & 57713.0 & 1057 & \phn8.4 & $0.636 \pm 0.026$ \\ 
\hline
\end{tabular}
\\
\tablenotetext{a}{The angular outer radius of a spherical shell model,
  fitted directly to the visibility measurements in the
  Fourier-transform (\uv) plane by least-squares.  The uncertainties
  consist of three parts, added in quadrature: the statistical
  uncertainty from the fit, 2\% of the geometric mean of the major and
  minor axes (an estimated jackknife uncertainty obtained by dropping
  one antenna, see text), and the scatter in fitted radii obtained
  when randomly varying the antenna gains by 10\%. In all cases, the
  statistical contribution is small compared to the other two.}
\end{minipage}
\end{table}

Taking the weighted average of the radius measurements at 8.4 and 22
GHz of $(6.40 \pm 0.26) \times 10^{16}$~cm (for $D = 15.1$~Mpc), we
find that the expansion speed between $t=0$~d and 384~d was $(19300
\pm 790) \cdot (D/[15.1\,{\rm Mpc}])$ Taking the differences between
adjacent pairs of our VLBI epochs, we can calculate three further
velocity values: $14500 \pm 3400$~\kms, $12100 \pm 4700$~\kms\ and
$14300 \pm 4200$~\kms\ at $t \simeq 514$~d, $\simeq 748$~d and $t
\simeq 955$~d, respectively, all for $D = 15.1$~Mpc.
At our last epoch at $t = 1057$~d, $\rx = (14.4 \pm 0.6)\cdot
(D/(15.1\,{\rm Mpc})$.

\subsection{Proper Motion}
\label{spm}

We take the fitted centre positions, obtained using only strictly
phase-referenced data, as the best estimate of the centre position of
SN~2014C at each epoch.  Doing a least-squares fit over the positions
of our four epoch results in proper motions of $-30, -12$ \muasyr\ in
RA and decl.\ respectively, with estimated uncertainties of 33
\muasyr.  For a distance of 15.1~Mpc, these proper motions translate
to velocities of $-2100, -900$~\kms\ in RA and decl.\ respectively,
with uncertainties of 2400~\kms.  To within our uncertainties, then,
the centre of SN~2014C is stationary.

\subsection{Parametrization of the Expansion Curve}
\label{sexpfit}

In order to parametrize the expansion curve using our measured radius
values, we fitted our measured values of \rx\ with two different
functions describing the evolution with time.  We fitted these
functions to our VLBI radius measurements using weighted least
squares.

This first function, which we call the ``powerlaw'' function, is an
uninterrupted powerlaw, $r = r_{\rm 1yr} (t \, /{\rm \, yr})^m$,
where $r$ is the radius of the supernova at time,
$t$, $r_{\rm 1yr}$ is the radius at $t = 1$~yr, and $m$ is the
powerlaw coefficient, often called the expansion parameter.  Such a
function is often used to describe the expansion of SNe, and many SNe
do indeed show a shock radius expanding approximately in this fashion,
with the best studied example being SN~1993J
\citep[e.g.][]{Marcaide+1997, SN93J-4, SN93J_Manchester}.
Such a powerlaw evolution of the radius is in fact suggested on
theoretical grounds.  In a supernova, generally a dual shock structure
is formed with a forward shock being driven into the CSM and a reverse
shock back into the ejecta.  It was shown that in the case that both
the ejecta and CSM densities are powerlaws in radius, a self-similar
evolution occurs, with the radius evolving as $r \propto t^m$ where
$m$ is in the range 0.6 to 1 \citep[e.g.][]{Chevalier1982b}, a model
often called the ``mini-shell'' model.

\begin{figure}
\centering
\includegraphics[width=\linewidth]{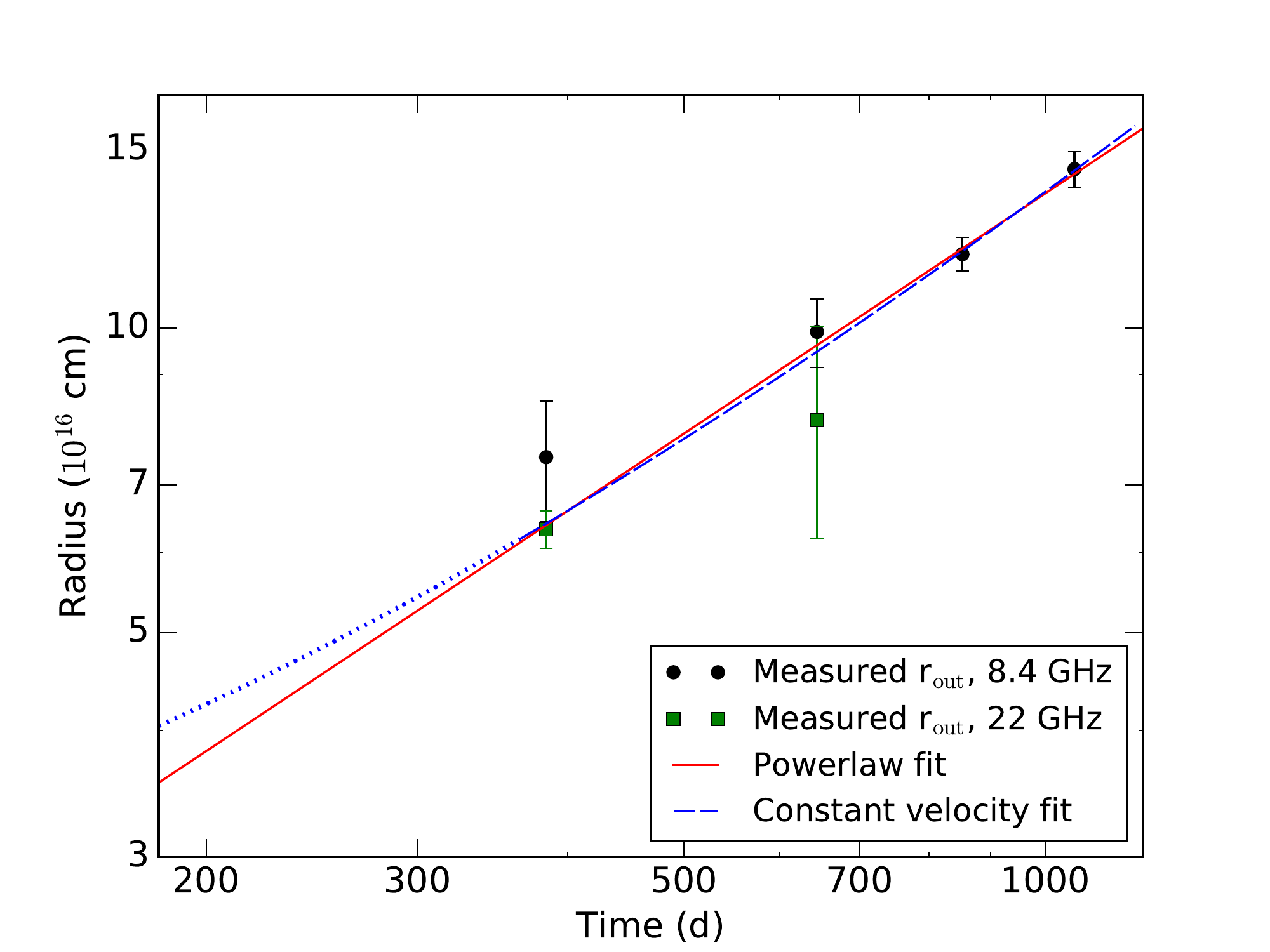}
\caption{The expansion of SN~2014C\@.  The outer radii were determined
  by fitting a spherical shell model directly to the visibilities (see
  Table~\ref{tradii}), and calculated for a distance of $D =
  15.1$~Mpc.  The squares (green) show the values at 22~GHz, and the
  circles (black) show the values at 8.4~GHz. The age is calculated
  assuming an explosion date of 2013 December 30\@. We show two
  different functions fitted to the measured radii.  The first, shown
  by the solid (red) line, is an uninterrupted powerlaw expansion of
  the form $r \propto t^m$.  The second, shown by the dashed (blue)
  line, is a constant velocity expansion after $t = 1$~yr (with an
  implied more rapid expansion before then).  We expect the
  approximately constant-velocity regime to begin at $t \sim 1$~yr,
  hence we show the extrapolation of the constant velocity fit to
  earlier times with a dotted line.  Note that in this logarithmic
  plot, the powerlaw expansion produces a straight line, while the
  constant velocity produces a curved line.}
\label{fexpand}
\end{figure}

Fitting such a function to our radius measurements, we obtain:
$$\rx = (6.16 \pm 0.19) \cdot \left(\frac{t}{\rm 1 \,
  yr}\right)^{(0.79 \pm 0.04)} \left(\frac{D}{15.1\,{\rm
    Mpc}}\right).$$ with of $\chi^2_4 = 2.19$, for
  four degrees of freedom.  This corresponds to a velocity of
$$v = (15355 \pm 910) \cdot \left(\frac{t}{\rm 1 \, yr}\right)^{(-0.21 \pm 0.04)}
  \, \left(\frac{D}{15.1\,{\rm Mpc}}\right) \; \kms.$$
We plot the fitted expansion curve as the red line Figure
\ref{fexpand}.

The fitted expansion curve, with $m = 0.79$, suggests a moderate
amount of deceleration.  This value is consistent with what is
generally expected from the mini-shell model.  If the CSM has a wind
density profile ($\rho \propto r^{-2}$), then the
mini-shell solution has that $m = (n - 3)/(n + 2)$
\citep{Chevalier1982b}, so this value of $m$ suggests ejecta with
$\rho \propto r^{-n}$ with $n = 6.9^{-0.8}_{+1.3}$.

However, SN~2014C is quite an unusual SN, and as already mentioned,
there strong evidence that the expanding shock encountered a region of
dense H-rich CSM at around $t \sim 130$~d.  The expected evolution in
this case differs from the self-similar powerlaw function
characterizing the evolution of the mini-shell model.  Similar systems
have been considered by, e.g., \citet{ChevalierL1989, ChugaiC2006,
  SmithM2007, vMarle+2010}.
At the point where the shock first encounters the dense shell, it
slows dramatically.  As the shock progresses through the dense shell,
it accelerates again due to the push from the ever-denser
undecelerated ejecta passing through the reverse shock.  Once the
whole of the massive CSM region has been shocked and accelerated, the
expansion continues at almost constant speed until the mass of the CSM
swept up from outside the massive shell becomes comparable to the
shell mass.  This behaviour has been reproduced in numerical
simulations by \citet{vMarle+2010}.

Since the shell impact for SN 2014C occurs before the time of the
first VLBI observations at $t = 384$~d, we can not resolve the period
of the first interaction with the CSM shell and the slowing of the
shock, so we model only the constant velocity expansion after the
impact of the shock on the massive shell. Therefore, the second
function that we fit to SN~2014C's expansion, which we call the
``constant velocity'' function, is $r [t > t\simpact] = r\simpact +
v\spost \cdot (t - t\simpact)$, where $t\simpact$ is
the time at which the shock impacts on the dense shell, $r\simpact$
is the radius at that time, and $v\spost$ is the shock velocity
after that time.  For $t\simpact \leq 1$~yr, that function is equal
to $r = r_{\rm 1yr} + v\spost \cdot (t - 1\;{\rm yr})$, so we fit the
latter function and avoid the problem of $t\simpact$ being not exactly
known.  We again fit the function to the VLBI radius measurements
using weighted least squares.\footnote{Note that the powerlaw function
  above also produces constant-velocity expansion when $m = 1$, but
  the constant velocity function differs from a power-law one with
  $m=1$ in that the intercept is fitted, in other words the
  extrapolated expansion curve of the constant velocity function is
  not forced to go through $r = 0$ at $t = 0$.}  We obtained
$$\rx = (6.19\pm0.19) + (4.29 \pm 0.19) \times
 \left(\frac{t}{\rm 1 \,
  yr} - 1\right) \left(\frac{D}{15.1\,{\rm Mpc}}\right).$$
where the fitted radius at 1 year is $(6.19\pm0.19) \times 10^{16}$~cm
and post-impact velocity is $v\spost = (4.29\pm0.19)\times
10^{16}$~cm~yr$^{-1}$, which is equal to $13600 \pm 650$~\kms.
The $\chi^2_4$ of this fit was 2.02, and we plot the
fitted function as the blue line in Figure~\ref{fexpand}.

The $\chi^2_4$ values for the powerlaw and the constant velocity
fitted functions were 2.19 and 2.02 respectively, and therefore our
data do not distinguish between these two functional forms, with the
constant velocity one providing an only insignificantly better fit.
The values of $\chi^2$ are close to the most probable values for four
degrees of freedom, indicating a reasonable fit.

\section{Discussion}
\label{sdiscuss}
\subsection{Morphology of SN 2014C}
\label{smorph}

The VLBI image of SN~2014C at $t = 1057$~d, when we had the highest
relative resolution, shows what appears to be a moderately circular
source, with a significant brightness enhancement or hot-spot to the W
side.  Except for this hot-spot, the image is approximately consistent
with the morphology expected from a spherically-expanding supernova,
which is that of a spherical shell in projection.  Such hotspots are
seen in the VLBI images of other SNe as well, for example in SN~1986J
\citep{SN86J-3}, SN~1993J \citep{SN93J-3} and SN~2011dh
\citep{SN2011dh_Alet}.  The cause of these asymmetries in the radio
brightness is not well known, but they are generally ascribed to
asymmetries in the density of the CSM or the ejecta.

\subsection{Radius and Expansion Speed}
\label{rradexp}

The VLBI measurements show that the radio emission region, the outer
edge of which is probably closely associated with the forward shock
\citep[see][for a discussion of this issue in the case of
  SN~1993J]{SN93J-4}, had a measured radius of $\rx = 6.40 \pm 0.26$
at $t = 384$~d, implying an average
expansion velocity up to that time of $19300 \pm 790$~\kms.  

By fitting the 16-GHz lightcurve, \citet{Anderson+2017} derive $\rx
\simeq 0.3$ at $t \simeq 80$~d, implying a velocity of
$\sim$5000~\kms.  Considerably higher velocities are suggested by
spectroscopy, with \citet{Milisavljevic+2015_SN2014C} finding
velocities of 13000~\kms\ (from Fe II, He I and Ca II absorption
lines) at $t = 10$~d, which suggesting that the forward shock must be
moving at very least at 13000~\kms.  In fact
\citet{Milisavljevic+2015_SN2014C} found evidence of a high-velocity
H$\alpha$ absorption feature at velocities as high as 21000~\kms\ at
$t = 10$~d.  Since the fastest H is difficult to detect in absorption,
the actual shock velocity is expected to be somewhat higher than
21000~\kms.  In the best-studied case of SN~1993J, the forward shock
velocity was $\sim$10\% higher than the blue edge of the H$\alpha$
absorption \citep{SN93J-4}, so we can estimate the shock velocity at
$t=10$~d as being $\gtrsim$23000~\kms\ from the H$\alpha$ absorption.

Our VLBI measurements of an average shock velocity of $19300 \pm
790$~\kms\ are in agreement with those estimated from optical
spectroscopy, but are incompatible with estimates of 
$\sim$5000~\kms\ from \citet{Anderson+2017}.
The estimates of size, and corresponding velocity, at $t \sim 80$~d
derived by \citet{Anderson+2017} are therefore clearly too small.
They were based on the assumption of SSA being the dominant absorption
mechanism for the first peak in the 16-GHz lightcurve.  Anderson et
al.\ did note the small velocities and questioned whether SSA was in
fact the dominant absorption mechanism.  They further suggested that
the second peak in the 16~GHz lightcurve is due to free-free
absorption by the dense CSM\@.  If that were the case, and if the
free-free absorbing material were spherically distributed, then there
must be significant free-free absorption even for the first lightcurve
peak, since the overlying dense CSM had at that point not yet been
shocked and was therefore still opaque.  Any sizes derived by assuming
predominately SSA, such as those of Anderson et al., will be therefore
be underestimated.

We can conclude that the shock front was likely expanding at
$\gtrsim$23000 \kms\ at $t = 10$~d.  Our powerlaw-function expansion
fit to the VLBI measurements after $t=384$~d, i.e.\ $\rx = (6.16 \pm
0.19) \; (t/{\rm 1yr})^{(0.79 \pm 0.04)}$, implies a forward shock
velocity at $t=10$~d of $33000 \pm 3200$~\kms, consistent with the
lower limit estimated from the H$\alpha$ absorption.

As already mentioned, however, given the strong indications from
optical, mid-infrared, X-ray, and radio that SN~2014C's shock
encountered dense, H-rich material sometime between a few weeks and
the first year after the explosion, an uninterrupted powerlaw
expansion seems unlikely.  The onset of the strong CSM interaction
must have happened sometime before $t = 130$~d, when strong H$\alpha$
lines were first observed \citep{Milisavljevic+2015_SN2014C}.

As we discussed in \S~\ref{sexpfit}, in this case an approximately
constant velocity expansion is expected after the shock has progressed
through the region of dense CSM\@.  Since our VLBI radius measurements
occur after the impact of the shock on the dense CSM, we fitted (in
addition to the powerlaw function), a constant-velocity function to
our VLBI radius measurements to accommodate this scenario.  The fit of
the two functional forms of the expansion is statistically
indistinguishable.  In other words, unfortunately, our VLBI
measurements do not shed much light on the evolution during the
critical period between $t = 30$~d and $\sim$1~yr when the shock first
encountered the dense H-rich material.

Regardless of whether we use the powerlaw or the constant-velocity
function for the expansion, however, there must have
been a rapid expansion up to the time of the first VLBI measurement at
$t = 384$~d, with the average expansion velocity to that time being
$19300 \pm 790$~\kms.  Since the velocity after $t=384$~d is
$\sim$30\% lower at $\sim 13600 \pm 650$~\kms, SN~2014C has been
substantially decelerated since the explosion.  If we extrapolate the
radius from the constant-velocity function backwards in time,
and match it to one computed from the 23000~\kms\ early velocity
derived from the H$\alpha$ absorption at $t=10$~d, we can calculate
that
the break in the expansion curve would have occurred at $t \sim
230$~d.  This is somewhat later than $t = 130$~d, when prominent
H$\alpha$ emission was already seen in the spectrum, suggesting
perhaps an even higher early velocity for the forward shock, or a
period of low velocity immediately after the impact, but before our
VLBI measurements, such as is seen in the simulations of
\citet{vMarle+2010}, see Figure~\ref{fvmarle} below.

Our size measurement of $\rx = 6.40 \pm 0.26$ 
at $t = 384$~d gives an upper limit on the outer radius of the dense
CSM shell, since our subsequent VLBI measurements show an
approximately constant velocity expansion, implying that the shock has
already exited the dense CSM shell.

A lower limit on the inner radius of the CSM shell can obtained by
extrapolating our constant velocity fit back to the time of impact,
$t\simpact$.  Lower values of $t\simpact$ lead to smaller values of
the inner radius of the shell.  If we take a low value of $t\simpact =
30$~d, we obtain $r_{\rm 16,\,extrapolated}\,(t=30~{\rm d})$ of $2.3
\pm 0.3$.  These values of $t\simpact$ and \rx\ imply an improbably
large pre-impact average speed of $87000 \pm 9900$~\kms. We therefore
consider somewhat later values of $t\simpact$ more likely, for
example, $t\simpact = 80$~d, which implies $r_{\rm
  16,\,extrapolated}\,(t=80~{\rm d})$ of $2.8 \pm 0.2$ and a
pre-impact speed of $41000 \pm 3500$~\kms.  We can therefore constrain
the dense CSM shell to be somewhere in the range of $2.8 \lesssim \,
\rx \, \lesssim 6.4$.  
This range of radii for the dense CSM shell is
consistent with the estimates given by
\citet{Milisavljevic+2015_SN2014C} and
\citet{Margutti+2017_SN2014C}.

These values are only approximate, since the exact evolution of the
shock radius around the time of the impact will be much more
complicated than an instantaneous transition from one constant
velocity to a lower one.

\subsection{Comparison to Numerical Models}
\label{svmarle}

Van Marle et al.\ (2010) \nocite{vMarle+2010} performed
simulations of a Type IIn SN shock hitting a dense circumstellar shell
that show in more detail the dynamics in this case.  We compare our
measurements of radius as a function of time to the evolution in those
simulations.  Our comparison is only intended to be illustrative, and
we scale the simulated velocity curves of \citealt{vMarle+2010} to the
case of SN~2014C in a very simple way which incorporates no physics.
Specifically, we first scale the times of the simulations so that the
impact of the shock on the dense CSM shell occurs at $t = 80$~d, and
then subsequently scale the simulation's velocities so that they match
the last one derived from our VLBI observations of SN~2014C, which was
14300~\kms\ at $t = 955$~d (\S \ref{ssize}).

The models of \citet{vMarle+2010} covered a wide range of parameter
space including CSM shell masses between 0.1 and 25 \Msol, supernova
explosion energies between $0.5 \times 10^{51}$ and $2 \times
10^{51}$~erg, stellar wind velocities of 50 to 500 \kms,
and mass-loss rates between $10^{-5}$ and $10^{-3}$ \Msolxyr.  The
progenitor of SN~2014C was likely a Wolf-Rayet star just before it
exploded, with wind velocity values towards the upper end, and
mass-loss rates towards the lower end, of those ranges.  Van Marle et
al.'s models were intended for a situation closer to that of $\eta$
Carinae, and therefore had dense CSM shells with inner radii of $(0.5
\sim 4) \times 10^{15}$~cm which were rather closer to the star at the
time of the SN than the one in SN~2014C, which is at $r \sim 6 \times
10^{16}$~cm.  \citep[For further details of the simulations, please
  see][]{vMarle+2010}.

\begin{figure}
\centering
\includegraphics[width=\linewidth]{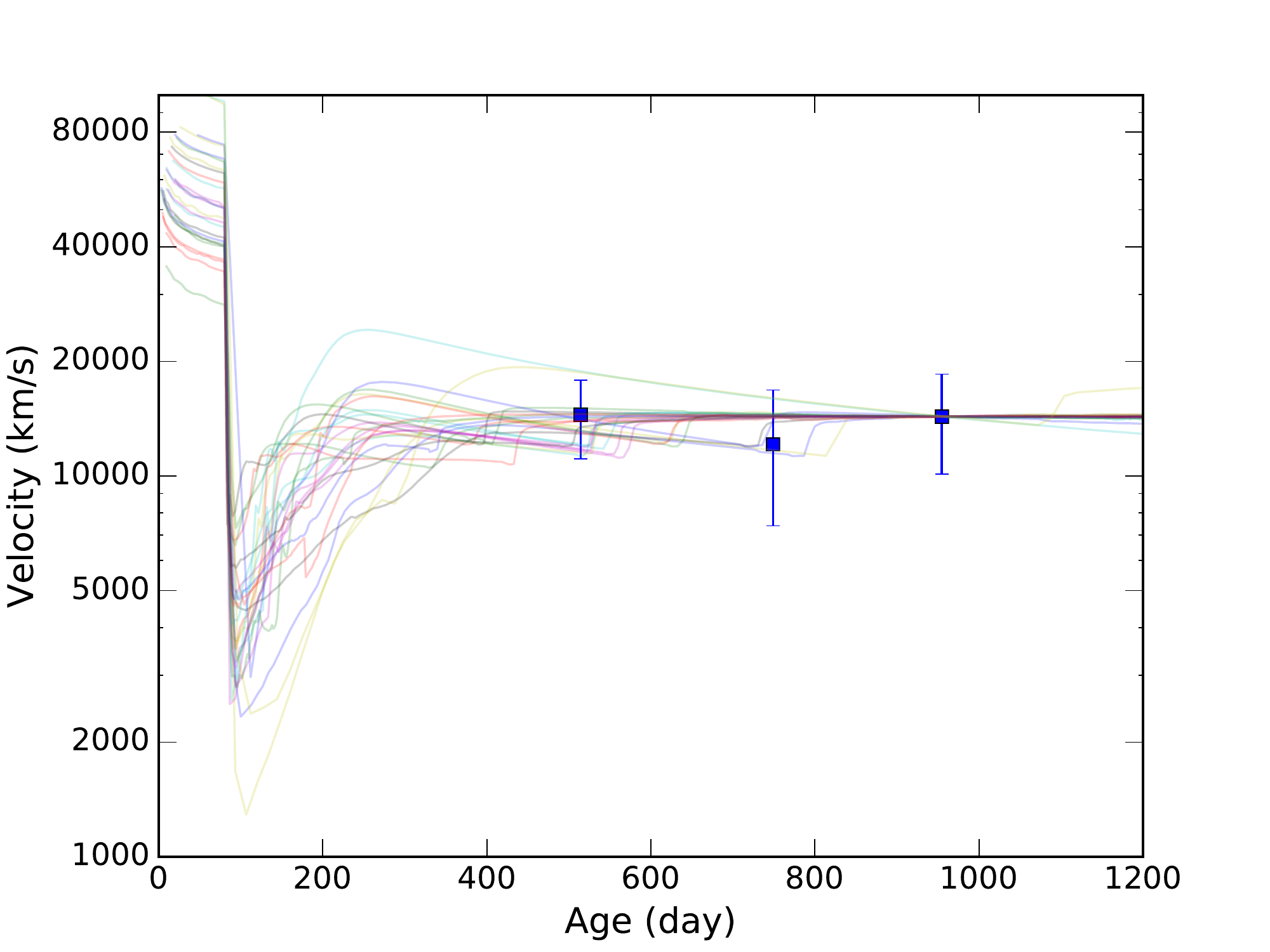}
\caption{The expansion velocity SN~2014C, compared to scaled
  simulations of a Type IIn supernova shock hitting a circumstellar
  shell from \citet{vMarle+2010}.  The three points show the
  velocities derived from pairs of our VLBI measurements of the outer
  radius of SN~2014C at four different times along with their
  uncertainties (see text, \S~\ref{ssize}).  The thin, coloured lines
  show all the simulations (``A00'', ``A01'', \dots ``G01''; for a
  range of different parameters for the explosion and the shell) from
  \citealt{vMarle+2010}, scaled in time so that the impact occurs at
  $t = 80$~d, and then scaled in velocity to match the measured value
  at $t = 954$~d, which was $14300 \pm 4200$~\kms. The figure is
  intended to be illustrative only since the scaling of the simulated
  velocity curves does not include any physics.}
\label{fvmarle}
\end{figure}

We plot the scaled, simulated velocity curves against our measurements
in Figure~\ref{fvmarle}.  Because of our simple scaling of the model
curves, the true evolution of SN~2014C is not expected to follow the
plotted curves in detail.  However, we believe that
Figure~\ref{fvmarle} is nonetheless useful to illustrate the general
nature of the behaviour.  It can be seen that in all the modelled
cases (when scaled to match SN~2014C), the initial velocities are
quite high, being in the range of $30000 \sim 80000$~\kms, and the
velocities then decrease very rapidly by factors of $4 \sim 8$ when
the shock impacts on the dense CSM shell, and then recover over the
next $\sim$200 days and remain relatively constant subsequently.  Thus
we expect in the case of SN~2014C that there would be very high shock
velocities in the first $\sim$80~d, before the shock impacted on the
dense CSM region, consistent with the $v \geq 21000$~\kms\ seen in
H$\alpha$ by \citet{Milisavljevic+2015_SN2014C}.  Indeed, shock speeds
of $\sim$20000 to $\sim$100,000 \kms\ are seen in other Type Ib SNe
\citep{Chevalier2007}, so a high (initial) shock speed is consistent
with what is expected.

\subsection{Radio Absorption and Ionization in the Dense CSM Shell}
\label{sabs}

The dense H-rich CSM must be at least partly transparent to radio
waves, otherwise the early radio emission, first detected at $t =
12$~d after shock breakout \citep{Kamble+2014b} and at $t = 17$~d at
15~GHz \citep{Anderson+2017} would have been absorbed.
If the dense CSM were distributed in a uniform spherical shell, with
mass $\sim$1~\Msol\ \citep{Margutti+2017_SN2014C}, then the densities
must be quite high, and strong free-free absorption would be expected
of the shell were ionized.  If we take the radius
$5.5\times10^{16}$~cm and thickness $10^{16}$~cm proposed by
\citet{Margutti+2017_SN2014C}, and assume a uniform density and
complete ionization, we can calculate a number density of electrons of
$n_e \sim 4\times 10^6$~cm$^{-3}$, which would have an optical depth
at 7~GHz of $\tau_{\rm 7GHz} > 100$. Our VLBI measurements constrain
the shell to be somewhere between the radii of $2.8\times10^{16}$~cm
to $6.4\times10^{16}$~cm (\S~\ref{rradexp}) at the time of impact.
Even if we take 1\Msol\ of ionized material uniformly spread out over
this whole range, we obtain $n_e \sim 10^6$~cm$^{-3}$ and $\tau_{\rm
  7GHz} \sim 90$.  \citet{Milisavljevic+2015_SN2014C} and
\citet{Margutti+2017_SN2014C} derive electron densities of a similar
magnitude from the optical spectra and X-ray emission respectively.
If the dense CSM shell were uniform and substantially ionized,
therefore, no radio emission should have been seen till the shock has
progressed through it and heated it to the point where it becomes
transparent to radio.

However, as discussed by \citet{vMarle+2010}, the densities in the
shell are likely high enough that substantial recombination would
occur, so it is possible that the dense CSM shell was mostly neutral
at the time of the SN explosion.  Nonetheless, the shock breakout will
almost certainly have at least partly ionized the dense CSM shell,
therefore a significant amount of free-free absorption of the radio
emission is likely to occur in the first year.  The exact amount of
free-free absorption will depend on the fraction of the CSM shell that
is ionized, and would be hard to calculate without detailed modelling.

The other possibility is that the dense CSM was not spherically
distributed \citep[note that][have already suggested this possibility
  for SN~2014C]{Anderson+2017}.  The CSM might be non-spherically
distributed on small scales, in other words fragmented.  Such
fragmentation of the CSM is often invoked in the case of Type II SNe
\citep[e.g.][]{Weiler+2002} to explain the relatively slow rise of the
lightcurves, so some fragmentation of SN~2014C's dense CSM shell is
not unexpected.  Alternatively, the dense CSM might be distributed
with some large-scale non-spherical geometry, perhaps in the form of
an equatorial disk.  Such a geometry is also not unexpected, since the
CSM structure in SN~1987A is seen to be quite complex \citep[see,
  e.g.][]{McCrayF2016}.  In the particular case of SN~2014C,
\citet{Milisavljevic+2015_SN2014C} discuss reasons why the mass loss
experienced by the progenitor system may have been strongly asymmetric
and how it was potentially driven by interaction between the
progenitor star and a close binary companion.

We can conclude that a dense CSM shell must be either largely neutral
or rather fragmented for the radio emission from the expanding shock,
detected already at $t = 12$~d, to escape before the shock has
progressed through the dense CSM shell.

\section{Summary and conclusions}

We obtained VLBI observations of the unusual supernova SN~2014C, which
allowed us to resolve the radio emission from the expanding shell
of ejecta, and directly measure the size at several epochs between $t
= 384$ and 1057~d after the explosion.  We found that:

\begin{trivlist}
  
\item{1.} At $t = 384$~d, the angular radius of the supernova was
  $0.283 \pm 0.012$~mas, corresponding to $(6.40 \pm 0.26) \times
  10^{16}$~cm (for a distance of 15.1~Mpc).

\item{2.} This radius corresponds to an average expansion speed up to
  $t=384$~d of $19300 \pm 790$~\kms.
\item{3.} Our VLBI measurements at $t \ge 384$~d show
  that the average speed between $t = 384$~d and 1057~d was $13600 \pm
  640$~\kms.  Our measurements are compatible with a constant velocity
  expansion during that period, but a modestly decelerating expansion
  over that period can not be excluded either.
\item{4.} The supernova has clearly been substantially decelerated
  already {\em before} our first VLBI measurement, that is, by $t =
  384$~d.
\item{5.} Our observations are consistent with the scenario, which has
  already been suggested based on observations at other wavelengths,
  that SN~2014C's expanding SN forward shock encountered a dense,
  H-rich CSM shell at {$30 \, \lesssim \, t \, \lesssim \, 130$~d}
  after the explosion.  An almost constant velocity expansion is
  expected after the shock emerges from the dense shell in this case,
  which is what we have seen in the VLBI measurements.


\item{6.} The detection of early radio emission before 30~d implies
  that the dense H-rich CSM must be largely transparent to radio.
  This could occur either if the dense CSM was mostly neutral or if it
  was non-spherically distributed, so that our lines of sight to near
  the explosion centre bypass most of the dense CSM.
  
\end{trivlist}

\section*{Acknowledgements }

We thank Allard Jan van Marle for making his simulation data available
to us.  We thank the anonymous referee for helpful comments.  We have
made use of NASA's Astrophysics Data System Abstract Service. This
research was supported by both the National Sciences and Engineering
Research Council of Canada and the National Research Foundation of
South Africa.

\bibliographystyle{mn2ev2} 
\bibliography{mybib1,sn2014c_temp}

\clearpage

\end{document}